\documentclass[letterpaper, 10 pt, conference]{ieeeconf} % Comment this line out if you need a4paper

\usepackage{mathtools}
\usepackage[listings,skins]{tcolorbox}

\usepackage{amssymb}

\usepackage{calc}
\usepackage{caption}
\usepackage{subcaption}

\usepackage[
  backend=bibtex,
  bibstyle=ieee,        % IEEE formatting for the reference list
  citestyle=numeric-comp, % compressed numeric citations: [1–3]
  defernumbers=true
]{biblatex}
\usepackage{multirow}
\usepackage{multicol}
\usepackage{amsthm}

\usepackage{colortbl}

\definecolor{gr}{rgb}{0.921, 0.972, 0.905}
\definecolor{pink}{rgb}{0.972, 0.905, 0.917}
\definecolor{redishh}{rgb}{0.9, 0.17, 0.31}
\definecolor{redish}{rgb}{1.0, 0.01, 0.24}
\definecolor{antique}{rgb}{0.57, 0.36, 0.51}
\definecolor{darkcandy}{rgb}{0.64, 0.0, 0.0}
\definecolor{pastel}{rgb}{0.09, 0.45, 0.27}

\IEEEoverridecommandlockouts        % This command is only needed if 
\title{\Large Fairness Definitions and Metrics in Deep Reinforcement Learning for Drug Discovery in Healthcare: A Rapid Evidence Review}

\author{\small \vspace{-3mm} Esmaeil Shakeri$^{1}$, Ronnie de Souza Santos$^{1}$, Behrouz Far$^{1}$% <-this  stops a space
\thanks{$^{1}$Esmaeil Shakeri, Ronnie de Souza Santos, and Behrouz Far are with the Department of Electrical and Software Engineering, Schulich School of Engineering, University of Calgary, Canada. 
  {\tt\scriptsize \{esmaeil.shakerihosse, ronnie.desouzasantos, far\}@ucalgary.ca}}
 }

\begin{document}
\sloppy
%\ninept
%
\maketitle
\thispagestyle{empty}
\pagestyle{empty}
\begin{abstract}

%%%%%%%%%%%%%%%%%%%%%%%%%%%%%%%%%%%%%%%%%%%%%%%%%%%%%%%%%

Deep reinforcement learning (DRL) is increasingly applied to de novo molecular design, but choices in data, rewards, and evaluation can yield uneven performance across disease areas and chemotypes. Despite this, there is no concise synthesis of how fairness is defined, measured, and tested in DRL-based drug discovery. In this rapid evidence review, we synthesize fairness definitions and metrics for DRL-driven molecule generation in healthcare. We focus on three questions: (i) how dataset composition and split strategies, especially scaffold versus random splits, affect evaluation and distribution shift; (ii) how reward design (e.g., QED, docking, toxicity, synthetic accessibility) can create or mitigate bias, with emphasis on cancer targets; and (iii) which measurable metrics best capture fairness. This includes parity across cancer versus non-cancer indications and across cancer subtypes. It also includes distributional balance in key physicochemical descriptors, scaffold/chemotype diversity, groupwise validity, toxicity, and synthetic accessibility. From 2017 onward, we searched major biomedical, computer science, and engineering literature databases and used arXiv for horizon scanning. Records were screened using PRISMA-style procedures and analyzed via content coding to link reported parity outcomes to dataset and reward choices. Our review provides a concise set of fairness definitions and metrics for DRL molecule generation. It offers practical guidance for reporting distribution parity and outcome parity. It also summarizes how dataset and reward choices relate to observed parity effects and identifies open gaps relevant to trustworthy, cancer-relevant DRL generation.

\end{abstract}

%%%%%%%%%%%%%%%%%%%%%%%%%%%%%%%%%%%%%%%%%%%%%%%%%%%%%%%%%

\section{INTRODUCTION}
\label{sec:intro}

DRL is increasingly used to generate and optimize small molecules by defining design as a sequential decision process, showing promise for de novo drug design and the exploration of the chemical space aimed at a specific goal \cite{popova2018deep, olivecrona2017molecular, shakeri2025accelerating}. Drug discovery itself is a multistage pipeline; researchers identify a biological target, screen and triage compounds, optimize leads for potency and safety, and then move to preclinical testing before human clinical trials \cite{hughes2011principles, paul2010improve}. Each stage narrows candidates and imposes practical constraints such as toxicity, synthesizability, and cost that shape molecules that advance \cite{paul2010improve, dimasi2016innovation}. These design and pipeline choices make the topic directly relevant to fairness definitions and metrics. Software fairness issues occur when a system’s built in biases unevenly impact certain groups, leading to ethically and socially harmful outcomes \cite{sotolani2025exploring, suresh2021framework}. In software fairness, a system is considered fair when it does not impose systematic, measurable disadvantages on defined groups \cite{mehrabi2021survey, friedler2019comparative,}. Fairness definitions and metrics operationalize this idea through criteria such as parity of selection or error rates, parity calibration, and distributional parity over representations such as scaffolds or physicochemical property bins \cite{barocas2023fairness, dwork2012fairness}. These criteria can conflict, so trade offs should be made explicit and transparently reported in high stakes healthcare contexts \cite{kleinberg2016inherent, mitchell2019model}. In DRL systems, data curation, reward functions, and modeling constraints can systematically favor certain chemotypes or disease areas, while overlooking candidates that score poorly on common drug-likeness proxies (e.g., molecular-weight thresholds, logP, and hydrogen-bond counts) \cite{chen2023algorithmic, ueda2024fairness, mitchell2019model, barocas2023fairness}. In this review, fairness means avoiding systematic disadvantage in meaningful groups in drug discovery. These groups include disease classes, chemotypes (e.g., Bemis–Murcko scaffolds) and constraints-related dimensions such as synthesizability and risk of toxicity \cite{bemis1996properties}. We also emphasize fairness reporting with clear definitions, metric selection, and calibration checks to support trustworthy decision making in DRL pipelines \cite{mitchell2019model, guo2017calibration, alvarsson2021predicting}. Rapid evidence review is needed because DRL models often optimize proxy rewards (e.g., drug likeness for molecular generation) that may not reflect social priorities or safety trade offs \cite{amodei2016concrete, popova2018deep, alvarsson2021predicting}, and the community lacks a concise synthesis of fairness definitions, metrics, and testing practices tailored to the generation of DRL molecules \cite{barocas2023fairness, brown2019guacamol}.

\textbf{Motivation:} Although DRL is already widely used for de novo molecular design, outcomes remain highly sensitive to choices that are often treated as technicalities. These include dataset selection, reward specification, and evaluation metrics. Prior benchmarking shows that corpus composition and task design can create apparent performance gains while masking scaffold or property level bias \cite{brown2019guacamol,cieplinski2023generative,renz2024diverse}. In clinical contexts, the fairness literature warns that such pipeline choices can systematically benefit certain groups, targets, or chemotypes \cite{yang2023algorithmic,smith2023bias}. Current practice lacks a shared fairness vocabulary, routine parity tests across indications and cancer subtypes, and transparent reporting of reward functions and dataset provenance. As a result, aggregate metrics (e.g., validity, QED) can appear strong while hiding skewed physicochemical profiles (logP, MW, HBD/HBA) or scaffold distributions \cite{cieplinski2023generative,renz2024diverse}. Against this backdrop, a rapid evidence review targeting fairness in DRL for drug discovery can consolidate practical definitions, diagnostic metrics, and testing procedures to detect/mitigate these skews \cite{ueda2024fairness}. This synthesis will help ensure that DQN workflow and similar DRL systems yield molecule sets that are high quality by conventional metrics and are transparently evaluated for balance across cancer vs non-cancer indications and among cancer subtypes. It will also support more data-driven and equitable strategies in modern drug discovery \cite{mashayekhi2024deep, vamathevan2019applications, walters2020assessing}.

\textbf{Contribution and Research Questions:} This paper delivers a concise, practice-oriented synthesis of fairness in DRL for de novo molecular design. We perform a rapid evidence review to map current practices, technical considerations, and known limitations. Accordingly, we address the following research questions (RQs):

\textbf{RQ1: How using and comparison of different dataset can make bias in the evaluation results for new molecular generation in drug discovery?}

We aim to explore how different datasets and split strategies especially scaffold vs random influence evaluation. We identify where distribution shift inflates performance. We also specify how to report parity by indication and scaffold to support more generalizable results.

\textbf{RQ2: How do different DRL reward designs and optimization strategies create or mitigate bias across disease areas, particularly cancer, in de novo molecular generation?}

We seek to identify how alternative DRL reward designs (e.g., QED, docking, toxicity, and synthesizability) create or mitigate bias across disease areas. Our focus is parity between cancer and non cancer targets.

\textbf{RQ3: Which metrics capture fairness in DRL-based molecule generation across disease areas, and do outputs show parity between cancer and non-cancer contexts in drug discovery?}

This RQ examines which metrics best capture fairness for disease areas, especially cancer. It includes assessing parity between cancer and non-cancer contexts and across cancer subtypes, as well as physicochemical parity for MW, logP, HBD and HBA bins. The analysis will also cover scaffold diversity, groupwise validity, toxicity, and synthetic accessibility score (SAS). Finally, it specifies how to report both the distribution parity and the outcome parity.

\section{BACKGROUND AND RELATED WORK}

%%%%%%%%%%%%%%%%%%%%%%%%%%%%%%%%%%%%
\textbf{DRL for Molecular Design:} DRL has established itself as a core technique for de novo molecular generation, where an agent learns to assemble molecules to maximize a reward function representing a desired property, such as Quantitative Estimate of Druglikeness (QED) or binding affinity \cite{popova2018deep, olivecrona2017molecular, brown2019guacamol}. The field has since evolved to incorporate multi-objective rewards that balance factors like SAS and toxicity \cite{cieplinski2023generative, renz2024diverse}. However, this focus on composite performance metrics often overlooks a critical issue: reward functions and training data can systematically skew the model’s output, favoring certain chemical profiles over others.

\textbf{Bias and Fairness in Algorithmic Systems:} In machine learning, fairness is typically assessed by measuring outcomes across meaningful groups to ensure that no group is systematically disadvantaged \cite{sotolani2025exploring, barocas2023fairness}. In healthcare AI, this translates to auditing models for performance disparities across patient demographics or disease subtypes \cite{chen2023algorithmic, yang2023algorithmic}. For molecular generation, the relevant "groups" are defined by chemical structures (e.g., Bemis-Murcko scaffolds \cite{bemis1996properties}) and therapeutic contexts (e.g., cancer vs. non-cancer targets). A DRL model trained on a biased dataset or with a narrow reward function may not generate viable candidates for underrepresented disease areas, effectively mistaking data or objective limitation for a chemical impossibility \cite{smith2023bias}.

\textbf{Synthesizing the Gap:} While benchmarks like Guacamol \cite{brown2019guacamol} evaluate overall performance, and the fairness literature provides measurement tools \cite{barocas2023fairness, mitchell2019model}, their intersection remains nascent. Current DRL practices lack standardized methods to diagnose and report biases arising from core pipeline components, namely dataset curation, data segmentation strategies, and reward design. This review directly addresses this gap by synthesizing evidence to build a practical framework for integrating fairness considerations into the evaluation and reporting of DRL-based molecular generators.
%%%%%%%%%%%%%%%%%%%%%%%%%%%%%%%%%%%%%

\section{Methodology}

\section*{III. METHODOLOGY: RAPID LITERATURE REVIEW}

\textbf{Scope.} To answer the RQs in Section~I, we conducted a Rapid Review (RR) of fairness in DRL for de novo molecular design. RRs provide timely, actionable syntheses by streamlining some steps of full systematic reviews while maintaining transparency and reproducibility \cite{garritty2021cochrane, tricco2017rapid}.

\subsection{Search Strategy}

\textbf{Search keywords} In each of the databases used in this review, advanced search filters were applied to restrict the results to articles published from January 2017 onward. For complete transparency, only full-text articles were manually reviewed. The search queries were formulated using Boolean logic as follows:

1. "Deep reinforcement learning" AND "drug discovery" AND "reward" AND "fairness" AND "healthcare".

2. "Deep reinforcement learning" AND "drug discovery" AND "QED" AND "parity" AND "cancer" AND "healthcare".

3. "Deep reinforcement learning" AND "drug discovery" AND "reward function" AND "reward shaping" AND "distribution shift" AND "healthcare".

4. "Deep reinforcement learning" AND "drug discovery" AND "dataset" AND "scaffold split" AND "random split" AND "distribution shift" AND "fairness" AND "healthcare".

Additionally, the OR operator was used to expand the search across these terms where appropriate.

\subsection{Databases}

We systematically searched major bibliographic databases and publisher platforms spanning healthcare, AI/ML, and drug discovery. Google Scholar was used for broad discovery and snowballing. PubMed targeted biomedical and clinical studies. IEEE Xplore and the ACM Digital Library covered technical AI/ML and systems research. SpringerLink, ScienceDirect, and other publisher platforms were used to capture multidisciplinary and applied work in computational chemistry and healthcare AI. We also searched journal collections from the Nature portfolio, JMIR, and ACS to ensure coverage of high-impact interdisciplinary and translational research. Finally, arXiv was screened to identify recent preprints in RL and responsible AI; however, preprints were not included in this rapid review.

\begin{figure}
  %\vspace{-4mm}
   \centering
  \includegraphics[width=\linewidth]{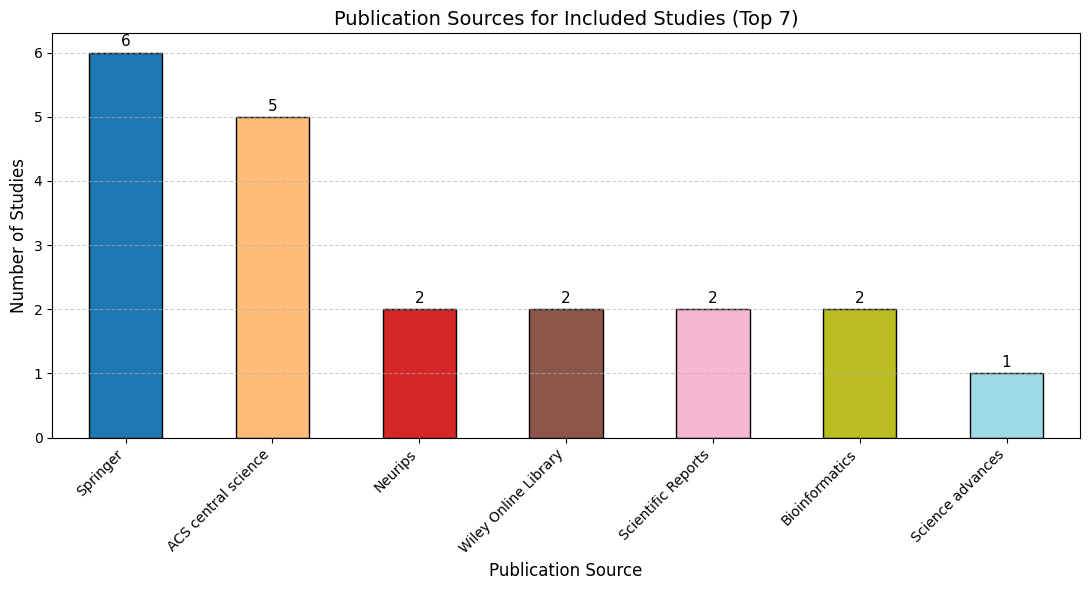}
   \caption{\footnotesize Distribution of included studies by publication source.}
     \label{fig:publicationsource}
  %  \vspace{-5mm}
     \end{figure}

To further characterize the dissemination venues of the included studies, we examined their distribution across publication sources. As shown in Fig.~\ref{fig:publicationsource}, the included studies are predominantly published by major scientific publishers. Springer represents the most frequent publication source, followed by ACS Central Science. Additional contributions appear in venues such as NeurIPS, Wiley Online Library, Scientific Reports, and Bioinformatics, while Science Advances is represented by a smaller number of publications. Overall, this distribution across the top 7 publication sources indicates that research in this area is primarily disseminated through well-established, high-impact journals and conferences, with a smaller presence across a diverse set of specialized venues.

\subsection{Inclusion and Exclusion criteria}

We used targeted filters to keep the results relevant and rigorous. The review focuses on recent work by DRL in healthcare, especially drug discovery.

%%%%%%%%%%%%%%%%%%%%%%%%%%%

\begin{itemize}
    \item \textbf{Inclusion--} Articles were included if they met all of the following:
    \begin{itemize}
        \item Written only in English.
        \item Published between January 2017 and October 2025.
        \item Available as peer-reviewed and full-text research (applied work or peer reviews).
        \item Focused on DRL for drug discovery with an emphasis on fairness metrics; other healthcare cases were included only when directly relevant.
    \end{itemize}

    \item \textbf{Exclusion--} For the exclusion criteria, we applied the following:
    \begin{itemize}
        \item Papers limited to theoretical or methodological discussion without an application in healthcare.
        \item Non-peer-reviewed materials (e.g., preprints, editorials, gray literature, and reports).
        \item Duplicate or secondary reports of the same study.
    \end{itemize}
\end{itemize}

This screening strategy yielded a focused corpus aligned with the aims of the study.

%%%%%%%%%%%%%%%%%%%%%%%%%%%

\subsection{Screening and selection process}

\begin{figure}
 % \vspace{-1mm}
   \centering
  \includegraphics[width=\linewidth]{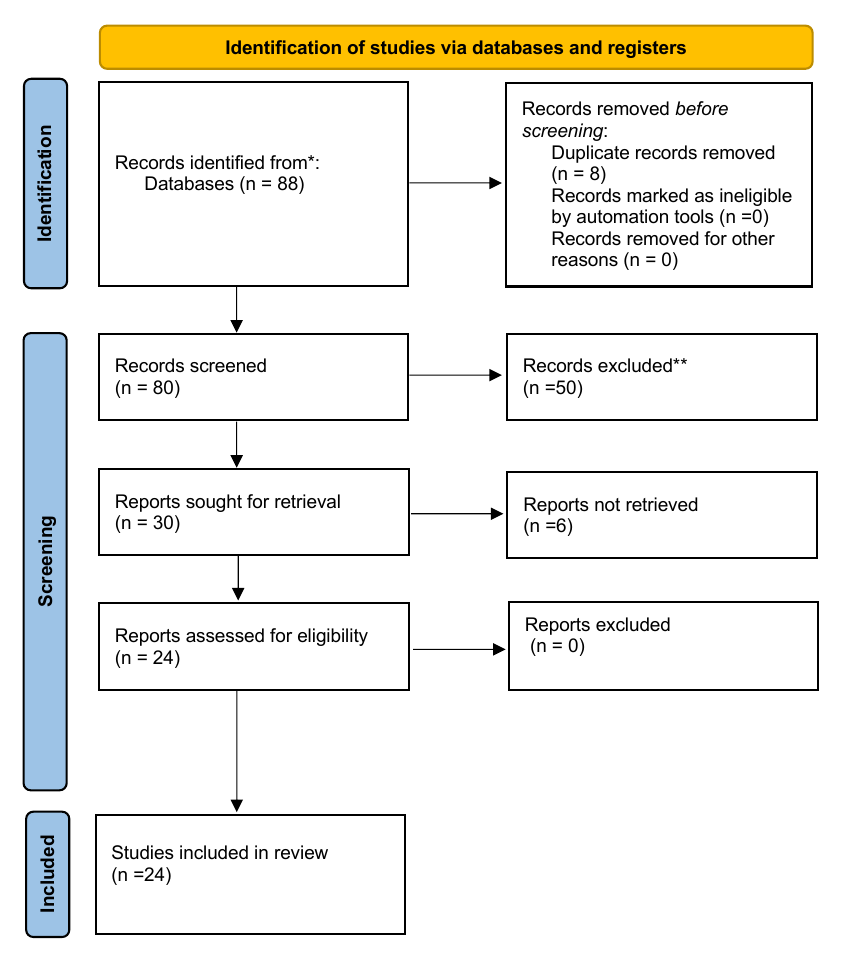}
   \caption{\footnotesize A PRISMA flow diagram illustrating the document screening and filtering process.}
     \label{fig:PRISMA}
   % \vspace{-4mm}
    \end{figure}

We followed a two-phase screening procedure consistent with PRISMA \cite{page2021prisma}. First, we identified 88 records from database searches, removed duplicate records ($n=8$), and screened the remaining titles and abstracts ($n=80$) against predefined inclusion and exclusion criteria, excluding 50 records. Next, we sought 30 reports for retrieval; 6 reports were not retrieved. We assessed 24 full-text reports for eligibility, and no full-text reports were excluded at this stage ($n=0$). To broaden coverage, we performed backward and forward citation tracking from key studies and added any additional records that met the criteria. The PRISMA flow diagram in Fig.~\ref{fig:PRISMA} summarizes the counts of records identified, screened and excluded, reports sought and not retrieved, full texts assessed, and included studies ($n=24$).

\subsection{Data Extraction}
We extracted short, relevant text snippets from each included source and stored them in a structured spreadsheet. This process focuses on our research questions and on DRL fairness definitions and metrics in healthcare drug discovery. A paper was included if it addresses at least one research question and is applicable to healthcare drug discovery; otherwise, we recorded the exclusion and reason. To improve reliability, a second reader reviewed the extracted units and the disagreements were resolved by discussion \cite{tricco2015scoping}. We report the final counts of the excluded and included studies.  As shown in Fig.~\ref{fig:PRISMA}, a total of 24 papers were included.

\begin{table}[t]
  \centering
  \scriptsize
  \caption{Primary studies identified during the review.}
 % \vspace{-1mm}
  \label{tab:primary-studies}
  \begin{tabular}{@{}lcccc@{}}
    \hline
    Source & Year & Methodology & Study type & RQ focus \\
    \hline
    Popova et al.             & 2018 & RL         & Applied & RQ2 \\
    Olivecrona et al.         & 2017 & RL         & Applied & RQ2 \\
    Yang et al.               & 2024 & RL         & Applied & RQ2 \\
    Wang et al.               & 2023 & RL         & Applied & RQ2 \\
    Bilodeau et al.           & 2022 & GAI        & Survey  & RQ1 \\
    Loeffler et al.           & 2024 & RL         & Other   & RQ2 \\
    Ciepliński et al.         & 2023 & GAI        & Other   & RQ1 \\
    Koziarski et al.          & 2024 & GAI        & Applied & RQ3 \\
    Zhou et al.               & 2019 & RL         & Applied & RQ1 \& RQ2 \\
    You et al.                & 2018 & RL         & Applied & RQ2 \& RQ3 \\
    He et al.                 & 2024 & RL         & Applied & RQ1 \& RQ2 \\
    Ståhl et al.              & 2019 & RL         & Applied & RQ2 \\
    Svensson et al.           & 2024 & RL         & Applied & RQ2 \\
    Bou et al.                & 2024 & RL         & Applied & RQ2 \\
    Hu et al.                 & 2025 & RL \& LLMs & Applied & RQ2 \\
    Shakeri et al.            & 2025 & RL         & Applied & RQ2 \\
    Liu et al.                & 2025 & RL         & Applied & RQ1--RQ3 \\
    Born et al.               & 2021 & RL         & Applied & RQ2 \\
    Mazuz et al.              & 2023 & RL \& LLMs & Applied & RQ2 \\
    Atance et al.             & 2024 & RL         & Applied & RQ2 \\
    Guo et al.                & 2025 & AI         & Other   & RQ2 \\
    Fang et al.               & 2023 & RL         & Applied & RQ2 \\
    Park et al.               & 2025 & RL         & Applied & RQ2 \& RQ3 \\
    Mercado et al.            & 2020 & RL         & Other   & RQ2 \\
    \hline
  \end{tabular}
 % \vspace{-4mm}
\end{table}

From this process, we retained 24 primary studies that met all eligibility criteria, and Table~\ref{tab:primary-studies} summarizes their key characteristics (publication type, methodological focus, and whether each study is applied, survey, or other). The review is centred on DRL; we only included non-Reinforcement Learning (RL) models when RL was used as a foundational optimization layer. We then applied a lightweight open-coding procedure to organize the studies under RQ1--RQ3, focusing on the dataset and split choice, reward design and optimization, evaluation and fairness metrics, and coverage of the disease area.

\begin{figure}
  %\vspace{-4mm}
   \centering
  \includegraphics[width=\linewidth]{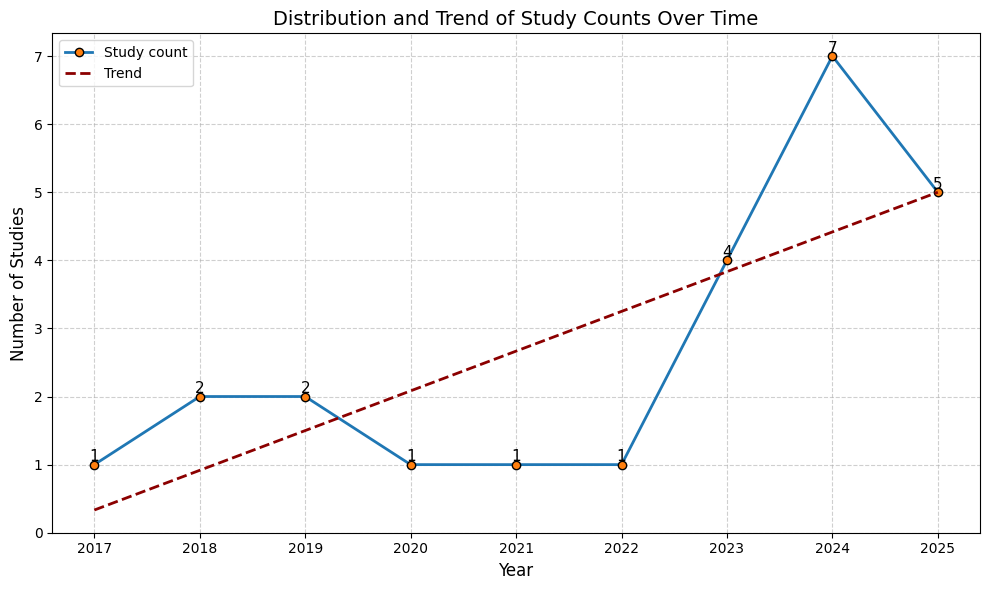}
   \caption{\footnotesize Distribution and Trend of Study Counts Over Time .}
     \label{fig:trend}
    %\vspace{-5mm}
     \end{figure}

Fig. \ref{fig:trend} presents the yearly distribution of the included studies over 2017–2025 and the corresponding fitted linear trend. Overall, the evidence base expands over time. The early period (2017–2019) shows limited and relatively stable output (1–2 studies per year), followed by a prolonged low-activity interval during 2020–2022 (approximately one study per year). In contrast, publication activity increases sharply from 2023 onward, rising to 4 studies in 2023 and reaching a maximum of 7 studies in 2024, before a modest decline to 5 studies in 2025. Although the counts fluctuate across years, the fitted trend line confirms a clear positive trajectory, indicating sustained growth in research attention and output in the later years of the study period.

\subsection{Geographic Distribution of Studies}

\begin{figure}
  %\vspace{-4mm}
   \centering
  \includegraphics[width=\linewidth]{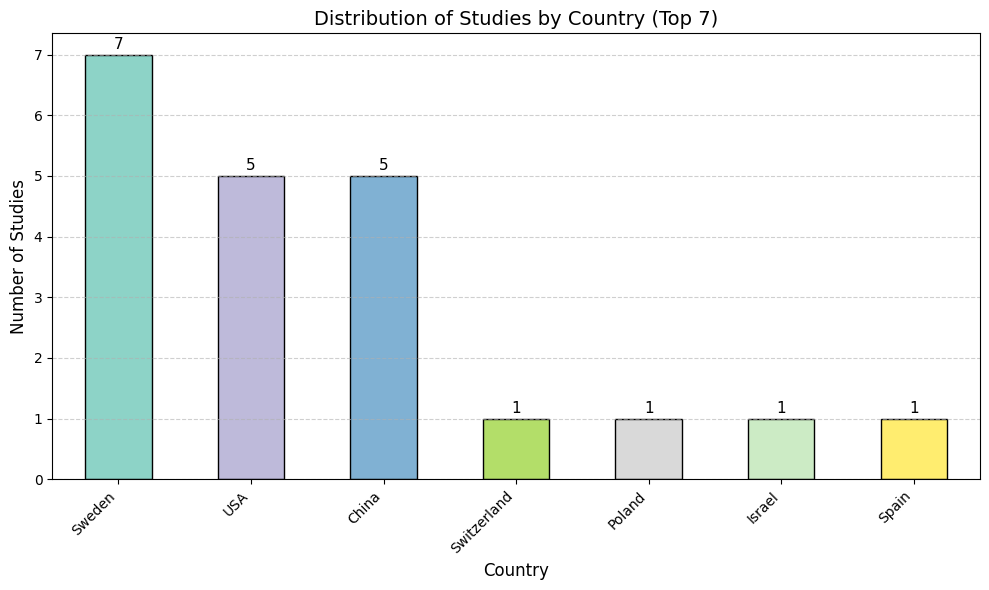}
   \caption{\footnotesize Distribution of the studies by the 7 most countries.}
     \label{fig:country}
   % \vspace{-5mm}
     \end{figure}

To complement the methodological and publication-source analyses, we also assessed the corresponding author’s country to characterize the geographic distribution of research activity. As shown in Fig.~\ref{fig:country}, presents the distribution of the included studies by country (top seven). Sweden contributes the highest number of studies, followed by the United States and China. Other countries, including Switzerland, Poland, Israel, and Spain, are represented by a smaller number of publications, indicating a more limited but geographically diverse research contribution.

\begin{figure}
  %\vspace{-4mm}
   \centering
  \includegraphics[width=\linewidth]{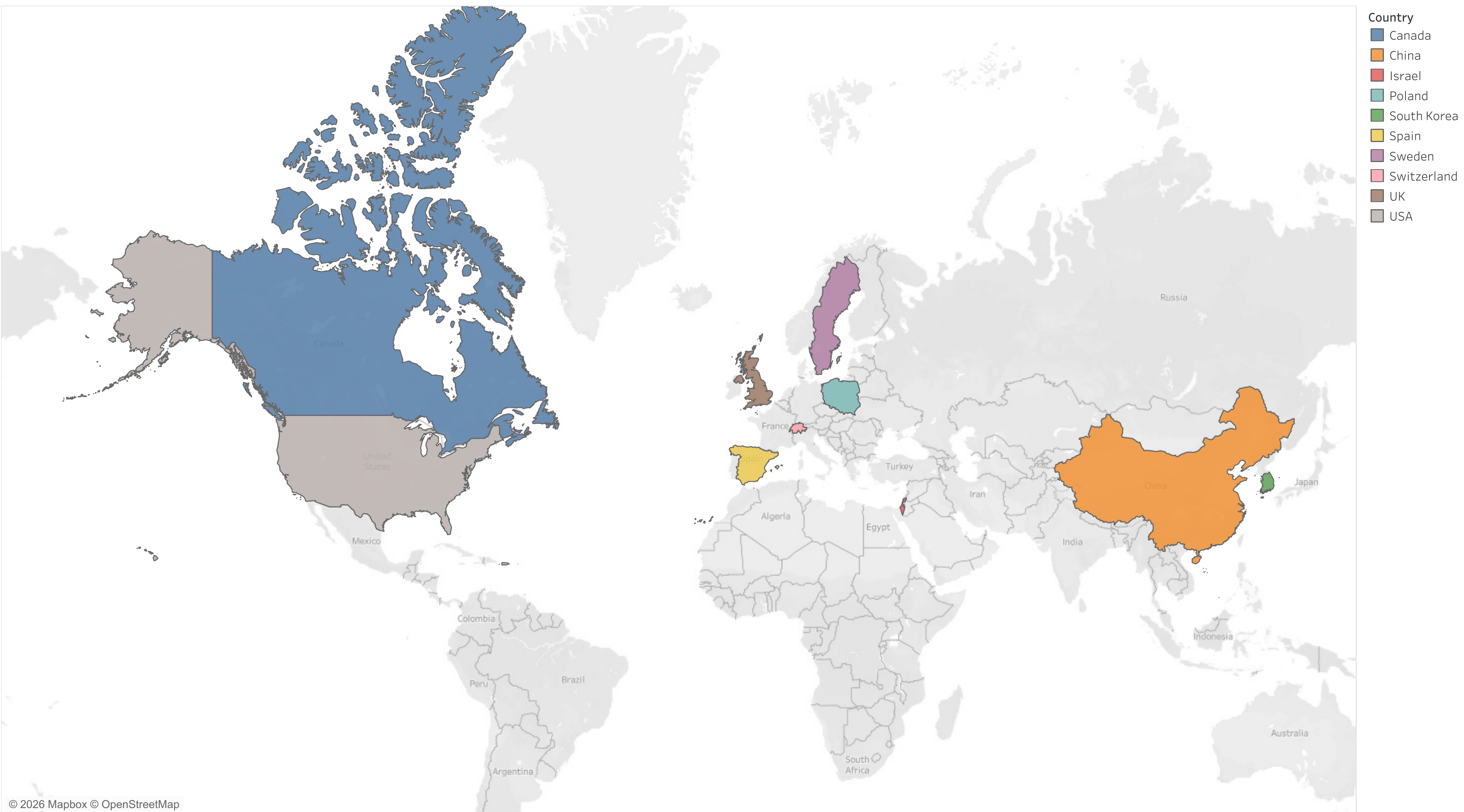}
   \caption{\footnotesize Map of the studies by countries.}
     \label{fig:demography}
    \vspace{-5mm}
     \end{figure}

Fig. \ref{fig:demography} provides a global overview of the geographic coverage of the included studies, with countries contributing at least one study highlighted on the world map. As illustrated, the research footprint spans North America, Europe, and Asia, demonstrating that the topic has attracted international attention. Contributions are geographically dispersed across multiple regions rather than confined to a single continent, with representation from both Western and Eastern research hubs. This spatial distribution indicates a broad global engagement with the research topic, while also suggesting that scholarly activity is concentrated in regions with established research infrastructure and active scientific communities.

\subsection{Types of Applications}

\begin{figure}
  %\vspace{-4mm}
   \centering
  \includegraphics[width=\linewidth]{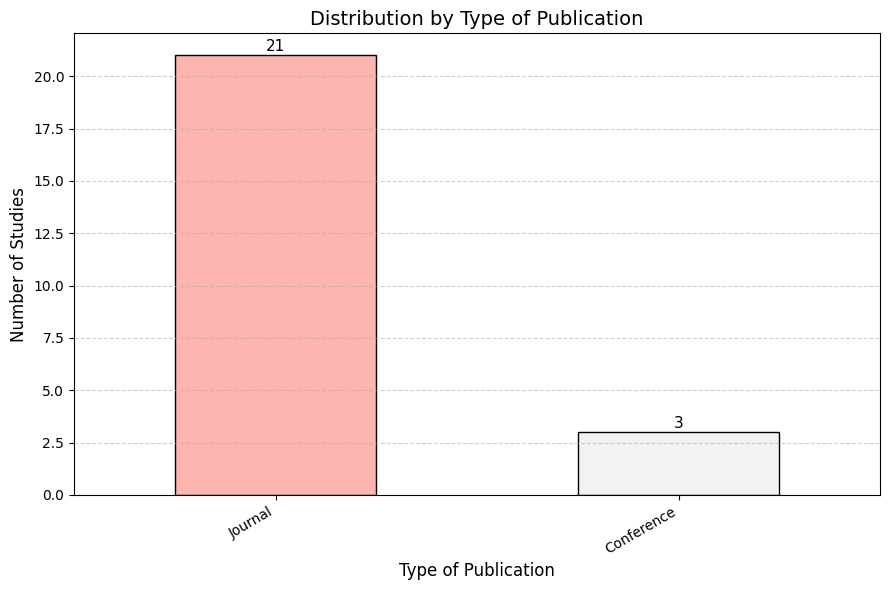}
   \caption{\footnotesize Distribution of type of publication.}
     \label{fig:publicationtype}
    \vspace{-5mm}
     \end{figure}

Fig. \ref{fig:publicationtype} illustrates the distribution of the included studies by type of publication. As shown, the literature is dominated by journal articles, which account for the vast majority of the studies (21 publications), while conference papers represent a much smaller portion (3 publications). This imbalance indicates that research in this area has primarily been disseminated through archival journal venues rather than conference proceedings, suggesting a focus on more mature, in-depth, and peer-reviewed contributions.

\subsection{Analysis Approach}

We first extracted short text segments (record units) from the included articles that directly address RQ1--RQ3. We then conducted a three-stage content analysis \cite{krippendorff2018content} with manual coding \cite{seaman1999qualitative} to identify issues and patterns related to datasets and splits, reward designs, and fairness definitions and outcomes of parity. In a pilot iteration, two studies were double-coded to draft and refine the initial codebook. A subsequent calibration iteration involved independent coding of a small set of studies to add the necessary tags such as calibration parity, scaffold diversity, groupwise validity, toxicity, and synthetic accessibility. This iteration was then used to finalize the coding guide. In the final iteration, one annotator coded all remaining units, and a second reviewer verified the codes, resolving disagreements through discussion. Beyond ensuring agreement, this multi-stage procedure was designed to support interpretive validity by forcing coders to discuss ambiguous cases and align on shared meanings of constructs such as parity, scaffold diversity, and fairness metrics, in line with RR guidance \cite{cartaxo2020rapid}. We analyzed the coded data by examining co-occurrence of determinants within each unit and aggregating counts across studies, while mapping fairness definitions and metrics to reporting practices. To ensure consistency, inter-rater agreement for screening was quantified with Cohen’s kappa \cite{mchugh2012interrater}, which indicated \emph{almost perfect} agreement between the reviewers ($\kappa = 1.00$). Coding reliability was also checked on a held-out subset using Cohen’s $\kappa$. The resulting outputs include a mapping of dataset and split choices to reward terms and parity outcomes, frequency visualizations, and a concise narrative of recurring patterns.

%%%%%%%%%%%%%%%%%%%%%%%%%%%%%%%%%%%%%%%%%%%%%%%%%%

\begin{table}[t]
  \centering
  \scriptsize
  \caption{Classification of the primary studies.}
 % \vspace{-1mm}
  \label{tab:classification}
  \begin{tabular}{@{}lll@{}}
    \hline
    Aspect & Approach & Sources \\
    \hline
    Datasets \& splits 
      & Benchmarks         & \cite{bilodeau2022generative,zhou2019optimization,liu2025diffmeta,gummesson2024utilizing} \\
      & Split strategy     & \cite{he2024evaluation,liu2025diffmeta,guo2025umap} \\
      & Docking benchmarks & \cite{cieplinski2023generative,he2024evaluation} \\
    \hline
    Reward design 
      & Property-based rewards   & \cite{popova2018deep,olivecrona2017molecular,you2018graph,zhou2019optimization,fang2023novo,staahl2019deep,wang2023molecular} \\
      & Disease-specific rewards & \cite{atance2022novo,mazuz2023molecule,born2021paccmannrl,yang2024enabling,liu2025diffmeta} \\
      & Bias-mitigating schemes  & \cite{fang2023novo,liu2025diffmeta,hu2025activity,bou2024acegen,park2025mol,loeffler2024reinvent,shakeri2025accelerating,mercado2020practical,wang2023molecular,staahl2019deep} \\
    \hline
    Fairness metrics 
      & Distribution-level       & \cite{liu2025diffmeta,koziarski2024rgfn} \\
      & Diversity \& synth.      & \cite{liu2025diffmeta,koziarski2024rgfn,you2018graph} \\
      & Outcome-level            & \cite{liu2025diffmeta,koziarski2024rgfn,you2018graph} \\
    \hline
  \end{tabular}
  %\vspace{-6mm}
\end{table}

%%%%%%%%%%%%%%%%%%%%%%%%%%%%%%%%%%%%%%%%%%%%%%%%%%

\section{RESULTS}

The RR was carried out in October 2025. The following subsections present the results of the knowledge synthesis, with a summary provided in Table \ref{tab:classification} All study materials, including the spreadsheets that contain detailed information on the collected papers and their mapping to the research questions, will be made available upon publication of this paper.

\subsection{Dataset choice, benchmarks, and split strategies (RQ1)}

For RQ1, we examined how dataset composition and splitting strategies bias the evaluation of DRL-based de novo molecular generation. Existing work spans benchmark platforms for unconstrained generation, task-specific activity datasets (e.g., DRD2 from ExCAPE-DB), and larger pretraining corpora derived from ChEMBL. Across these settings, there is substantial variation in how distribution shift and scaffold-level generalization are handled \cite{gummesson2024utilizing}.

\noindent\textbf{1) Benchmark composition and distributional bias:}
Bilodeau et al.\ review how standard generative benchmarks based on ZINC and ChEMBL drug-like subsets define validity, novelty, and distributional fidelity metrics while covering only a limited portion of the broader molecular discovery space~\cite{bilodeau2022generative}. These benchmarks encourage models to match a relatively narrow chemical prior and do not explicitly encode a therapeutic indication. Molecule Deep Q-Networks (MolDQN) is evaluated on such proxy tasks and synthetic optimization benchmarks and explicitly notes that many of these tasks are not representative of real optimization problems in drug discovery~\cite{zhou2019optimization}. DiffMeta-RL pretrains on a filtered ChEMBL corpus while benchmarking on curated evaluation sets. It explicitly removes exact SMILES overlaps between pretraining and benchmark molecules to prevent data leakage, so that generative metrics reflect genuine generalization rather than memorization~\cite{liu2025diffmeta}. Together, these choices illustrate how benchmark design can either inflate performance when leakage or overly aligned distributions are present. In contrast, explicit deduplication and filtering provide a more realistic assessment of model performance.

\medskip
\noindent\textbf{2) Split strategies and scaffold-level generalization:}
In all included studies, random splits remain the dominant strategy for supervised components, such as activity prediction models. For example, the Dopamine Receptor D2 (DRD2) activity model used to guide transformer-based RL generation in He et al.\ is trained on the ExCAPE-DB dataset with a conventional random partition into training, validation, and test sets~\cite{he2024evaluation}. This setup evaluates performance under relatively mild distribution shift, because analogs and closely related scaffolds can appear across partitions. DiffMeta-RL, by contrast, introduces both random and scaffold-based splits for the CYP450 inhibition dataset, using Bemis–Murcko scaffolds to stratify molecules and explicitly reduce scaffold overlap between training and test sets~\cite{liu2025diffmeta}. The authors highlight the scaffold split as a more challenging and realistic benchmark that probes generalization to unseen scaffolds rather than to close analogs \cite{guo2025umap}.

\medskip
\noindent\textbf{3) Dataset realism and indication coverage:}
Several studies argue that widely used proxy benchmarks, including unconstrained logP optimization and QED optimization on generic datasets, fail to reflect the complexity of real discovery problems. MolDQN notes that popular optimization tasks are often overly simplified and do not align with the optimization of medicinal chemistry lead~\cite{zhou2019optimization}. Ciepli'{n}ski et al.~\cite{cieplinski2023generative} propose a docking-based benchmark in which generative models are evaluated using SMINA docking scores on realistic protein targets~\cite{koes2013lessons}. Despite this more realistic setting, generative methods rarely outperform the top 10\% docking scores observed in the ZINC training or reference sets. This gap highlights how sensitive the reported performance remains to limited, target-specific datasets. Transformer-based RL experiments for DRD2 operate similarly within a narrow indication and scaffold regime, selecting starting compounds from DRD2 actives in ExCAPE-DB and focusing evaluation entirely on this single target~\cite{he2024evaluation}. None of these works stratifies benchmark performance by therapeutic area (e.g., cancer vs.\ non-cancer) or reports parity across targets, even though the underlying datasets are strongly skewed toward particular indications such as CNS receptors.

\subsection{Reward design and optimization strategies (RQ2)}

\noindent\textbf{1) Property-based reward functions:}
In all included studies, DRL for de novo molecular generation is predominantly framed as a property or target–optimization problem, with the disease area being only indirectly specified through surrogate models and benchmarks. Most works optimize generic proxies such as QED, penalized logP, and SAS, sometimes combined into composite scores (e.g., QAscore), rather than encoding disease–specific endpoints. Graph Convolutional Policy Network (GCPN)~\cite{you2018graph} and MolDQN~\cite{zhou2019optimization} are canonical examples, using single- or few-objective rewards based on penalized logP and QED, while Popova et al.~\cite{popova2018deep}, Olivecrona et al.~\cite{olivecrona2017molecular}, and the quality assessment-based drug design approach (QADD)~\cite{fang2023novo} extend this pattern with SMILES-based and quality-score-based optimization \cite{staahl2019deep, wang2023molecular}. Strong optimization pressure on simple proxy objectives can push models toward extreme and often unrealistic chemotypes. For example, when penalized logP is optimized without explicit constraints, the agent tends to build oversized, highly lipophilic structures; MolDQN~\cite{zhou2019optimization} shows that unconstrained logP maximization can effectively collapse into a trivial “add–carbon” policy. Constraints and adversarial components are therefore often needed to keep the search in realistic regions: GCPN~\cite{you2018graph} combines property scores with chemical validity rules and similarity penalties, while QADD~\cite{fang2023novo} iteratively retrains its quality–assessment model on low–quality RL outputs to progressively shift the reward landscape towards more drug–like molecules.

\medskip
\noindent\textbf{2) Disease-specific reward functions:}
The choice of scoring functions strongly shapes which disease areas are effectively represented. A substantial subset of studies uses DRD2 (or related dopamine receptors) as the primary activity benchmark, such as in MolDQN~\cite{zhou2019optimization}, GCPN~\cite{you2018graph}, and subsequent graph and SMILES-based generators~\cite{atance2022novo, mazuz2023molecule}. Loeffler et al.~\cite{loeffler2024reinvent} show that relying only on top–scoring molecules concentrates the policy on narrow chemotypes, while replaying a mixture of high, intermediate and low–scoring compounds improves diversity at the cost of slower convergence. RL fine-tuning of graph-based models such as GraphINVENT~\cite{atance2022novo} similarly optimizes activity on popular central nervous system (CNS)–related benchmarks, reinforcing a focus on dopamine–receptor targets. Oncology-focused work illustrates another form of bias introduced by biologically grounded rewards. PaccMannRL~\cite{born2021paccmannrl} conditions the generation of molecules on cancer gene-expression profiles and uses an anticancer drug-sensitivity model as the primary reward, optionally including toxicity predictors from the Tox21 and SIDER datasets. Together, these components drive the model to propose hit-like molecules tailored to specific cancer cell lines and cancer types~\cite{born2021paccmannrl}. Related pipelines optimize docking or kinase–inhibitor activity within narrow oncology–oriented panels~\cite{yang2024enabling, liu2025diffmeta}. Across these studies, disease-aware rewards are therefore concentrated almost entirely on cancer targets, while non–cancer indications remain underexplored.

\medskip
\noindent\textbf{3) Bias-mitigating strategies:}
Several recent DRL frameworks explicitly move beyond simple property proxies toward multi-objective and ADMET-aware reward designs, where ADMET refers to absorption, distribution, metabolism, excretion, and toxicity. QADD~\cite{fang2023novo} combines a learned QAscore with QED and SAS to guide the agent toward molecules that jointly satisfy potency, drug-likeness, and synthesizability. DiffMeta-RL~\cite{liu2025diffmeta} integrates a graph diffusion generator with RL and a cytochrome P450 (CYP450) inhibition predictor (MetaCYP), yielding compounds with improved metabolic stability and reduced CYP2C19/CYP3A4 liability. ACARL~\cite{hu2025activity} incorporates activity–cliff–aware rewards and contrastive objectives to preserve structure–activity cliffs across multiple targets, discouraging overly smooth activity landscapes and promoting the retention of high–affinity chemotypes. 
The optimization strategy also interacts with the reward design. MolDQN~\cite{zhou2019optimization} demonstrates that greedy and unconstrained maximization of penalized logP leads to unrealistic chemotypes, whereas carefully tuned $\epsilon$–greedy exploration and bootstrapped DQN improve exploration. Loeffler et al.~\cite{loeffler2024reinvent} report that replay buffers that mix low- and high–reward trajectories increase scaffold diversity and the number of predicted actives, suggesting that replay design can counteract reward-driven mode collapse. Toolkits such as ACEGEN~\cite{bou2024acegen} standardize policy gradient and actor–critic implementations, while curiosity-driven designs such as Mol-AIR~\cite{park2025mol} encourage the exploration of novel chemotypes beyond narrow high-reward basins. Increasing the action space in DQN–based generators (e.g., richer SMILES or graph edits) can further expand the chemical-space coverage without sacrificing validity when combined with penalties for unrealistic molecules~\cite{shakeri2025accelerating, mercado2020practical, staahl2019deep, wang2023molecular}.

\subsection{Fairness metrics and disease--area parity (RQ3)}

In addressing RQ3, we analyzed how the metrics already reported in DRL-based molecule generation studies can be interpreted through a fairness lens across disease areas. Rather than introducing new fairness-specific scores, the included works rely on existing evaluation families that can be repurposed for this goal: (i) distribution-level descriptors of physicochemical space, (ii) structural diversity and groupwise validity, and (iii) outcome-level performance indicators that could, in principle, be stratified by disease group.

\noindent\textbf{1) Distribution-level metrics as proxies for parity:}
All three included studies report standard generative or optimization metrics, which naturally extend to fairness analysis. DiffMeta-RL assesses the validity, uniqueness, novelty, internal diversity, and property distributions of generated molecules. It also compares key descriptors such as logP, QED, SAS, and topological polar surface area (TPSA) between generated and reference sets~\cite{liu2025diffmeta}. These quantities can be reused as distributional parity metrics by stratifying MW, logP, HBD, and HBA into bins and comparing their histograms across disease groups, for example between cancer and non-cancer targets or among different cancer subtypes. Similarly, Reaction-GFlowNet (RGFN) reports distributions and summary statistics of MW, QED, and SAS for the top-reward “modes” identified by the model~\cite{koziarski2024rgfn}. When reported per task or per oracle, these descriptors provide a direct handle on physicochemical parity across indication groups.

\medskip
\noindent\textbf{2) Structural diversity, synthesizability, and groupwise validity:}
Structural fairness is partially captured through scaffold and mode diversity. DiffMeta-RL analyzes scaffold recovery, ring-system novelty, and atom/bond-type distributions relative to a curated ChEMBL dataset and includes PAINS filters and charge-neutrality checks to quantify chemical realism~\cite{liu2025diffmeta}. RGFN operates directly in a reaction and building-block space and evaluates synthesizability through metrics such as SAS and AiZynthFinder scores, alongside reward distributions and counts of high-reward, low-similarity modes for each task~\cite{koziarski2024rgfn}. Together, these metrics can be organized as groupwise validity and diversity indicators: for each disease area, one could report (i) the fraction of valid and synthesizable molecules, (ii) scaffold or mode diversity, and (iii) distributions of SAS and toxicity flags. In contrast, GCPN primarily reports improvements in single-property objectives such as penalized logP and QED and in constrained optimization benchmarks~\cite{you2018graph}. It does not provide explicit groupwise validity or diversity analysis by indication. Although none of the three studies explicitly stratifies these structural metrics by disease area, their existing protocols could easily be extended to enable disease-specific analysis.

\medskip
\noindent\textbf{3) Outcome parity and current reporting gaps:}
Outcome metrics such as docking scores, predicted activity, or proxy rewards are reported per task or oracle, but not explicitly analyzed for parity between cancer and non-cancer indications. DiffMeta-RL demonstrates multiobjective improvements in both metabolic stability, via CYP450 inhibition predictions, and docking affinity for proton pump inhibitors, tracking reward curves, QED, and SAS over optimization rounds~\cite{liu2025diffmeta}. RGFN compares synthesizability-aware objectives and high-reward mode counts across docking and senolytic tasks~\cite{koziarski2024rgfn}. GCPN focuses on goal-directed optimization of penalized logP, QED, and similarity-constrained objectives, measuring success rates and property distributions on standard benchmarks~\cite{you2018graph}. However, none of these works reports or analyzes their outcome metrics such as hit rate, activity distributions, or docking score distributions by disease group or cancer subtype, and none explicitly introduces fairness or parity metrics.

\section{DISCUSSION}

This review shows that fairness in DRL-based molecular generation is largely shaped by three design choices: (i) dataset composition and split strategy, (ii) reward design and optimization, and (iii) the selection and reporting of evaluation metrics. Across RQ1 to RQ3, current practice appears optimized for generative quality and task performance rather than for parity across disease areas or chemotypes.

For RQ1, our synthesis indicates that widely used benchmarks and datasets encode narrow chemical priors. They also contain strong indication-specific biases. Drug-like ZINC and ChEMBL subsets and single-target datasets such as ExCAPE-DB DRD2 or CYP450 panels are convenient for method development. However, they rarely support scaffold-based evaluation. They also rarely enable stratified reporting by indication. Gains on these proxy benchmarks may not transfer to realistic discovery settings. In addition, the lack of scaffold-based splits and indication-aware reporting directly weakens any claim of fairness or parity in model behavior.

For RQ2, we find that reward functions are usually based on simple property proxies such as QED, logP, and SAS. They are also often designed around a small set of widely used targets. Disease-specific rewards are concentrated almost entirely in oncology. This creates a risk of reinforcing cancer-centric bias. For example, oncology-focused, transcriptomics-based rewards, such as those used in PaccMannRL, may deepen this imbalance. In parallel, dopamine receptor benchmarks can reinforce a CNS-centric chemical space. Multi-objective and ADMET-aware designs, together with more advanced exploration strategies, move closer to development-relevant goals. However, they are not yet evaluated for parity across different therapeutic areas.

For RQ3, current DRL studies already report many metrics that could function as fairness indicators if they were stratified by group. These metrics include physicochemical distributions such as MW, logP, and HBD/HBA bins. They also include scaffold or mode diversity, groupwise validity, synthesizability, and ADMET-related properties. However, these metrics are rarely used in a disease-aware way. Distribution parity in MW, logP, and HBD/HBA bins is typically not reported. Outcome parity is also missing, including hit rates and docking score distributions across cancer and non-cancer indications and across cancer subtypes. As a result, disparities can remain hidden behind aggregate metrics such as validity or QED.

Taken together, these findings suggest concrete directions for fairer DRL workflows in drug discovery, including our own DQN-based pipeline. Dataset and split choices should be documented and evaluated for scaffold-level generalization and indication coverage. Reward functions should treat disease area as an explicit design dimension rather than an implicit side effect of data availability. Evaluation protocols should routinely stratify existing metrics by disease group and chemotype and report both distribution and outcome parity. These steps would not change the underlying algorithms but would make fairness considerations visible and actionable, supporting more trustworthy and cancer-relevant DRL-based molecular generation.

\subsection{Research Implications}

\subsubsection{Implications for Research}
The synthesis presented in this rapid review implies that advancing fairness in DRL-based de novo molecular design requires a shift in how research problems are formulated, evaluated, and reported, rather than further emphasis on model architecture alone. The reviewed evidence collectively indicates that fairness-related behaviors emerge from interactions among dataset composition, split strategy, reward formulation, and evaluation protocol, suggesting that these elements should be treated as explicit research variables.
From a research-design perspective, future studies should prioritize evaluation settings that reflect chemically and clinically realistic generalization. In particular, scaffold-based splits and indication-aware benchmarks should be adopted more systematically to expose distribution shifts that remain hidden under random splits. Without such evaluation regimes, claims of robustness or generalization risk being overstated, especially for cancer-relevant discovery tasks where chemical diversity and transferability are central concerns.
The findings further imply that fairness should be conceptualized as a multi-dimensional construct in molecular generation research. Rather than focusing solely on aggregate success rates or reward maximization, future work should examine outcome parity, distributional balance in physicochemical properties, scaffold and chemotype diversity, and groupwise validity across disease areas. The absence of standardized fairness definitions and reporting practices highlights an opportunity for the community to develop shared benchmarks, metric taxonomies, and reporting guidelines that enable meaningful comparison across studies and therapeutic domains.

\subsubsection{Implications for Practice}
For practitioners, including machine learning engineers and pharmacists applying DRL-based molecular generation in real-world discovery pipelines, the results of this review suggest concrete changes to model development and validation workflows. Existing evaluation metrics routinely reported in DRL studies can be repurposed to assess fairness by stratifying them across disease indications, cancer subtypes, and chemotype groups, rather than aggregating results across heterogeneous tasks.
Practitioners should also recognize reward design as a central lever that shapes the effective coverage of chemical and therapeutic space. Choices regarding the weighting of docking scores, quantitative drug-likeness, toxicity predictions, and synthetic accessibility implicitly prioritize certain molecular subpopulations while suppressing others. Auditing reward functions, replay strategies, and dataset splits during development can help identify unintended biases early and reduce the risk of narrow chemotype exploitation, particularly in oncology-focused applications.
Overall, these implications indicate that fairness-aware practices can be incorporated into existing DRL pipelines without altering underlying algorithms. By documenting dataset assumptions, adopting more realistic validation strategies, and routinely reporting stratified evaluation metrics, practitioners can make fairness effects visible, interpretable, and actionable. This supports more trustworthy and clinically relevant use of DRL-based molecular generation in early-stage drug discovery.

%%%%%%%%%%%%%%%%%%%%%%%%%%%%%%%%%%%%%%%%%%%%%%%%%%
\section{STUDY LIMITATIONS}

This RR queried a selected set of databases (Google Scholar, PubMed, IEEE Xplore, ACM Digital Library, SpringerLink, Elsevier, ScienceDirect, Nature portfolio, JMIR, ACS, Science Advances) and limited inclusion to studies published from 2017 onward while excluding arXiv preprints. This coverage choice is a first limitation, as it may reduce recall and omit relevant earlier or non-indexed work. We mitigated this through forward and backward citation chaining and broad discovery in Google Scholar. Screening and record-unit coding were primarily performed by one reviewer. This constitutes a second limitation, because it introduces potential selection and interpretation bias. A second reviewer verified decisions, disagreements were resolved by consensus, and inter-rater agreement was reported using Cohen’s kappa.
%%%%%%%%%%%%%%%%%%%%%%%%%%%%%%%%%%%%%%%%%%%%%%%%%%

\section{CONCLUSION AND FUTURE WORK}

All in all, fairness-aware DRL for molecular design is a promising way to make de novo drug discovery more systematic, transparent, and data-driven. Our RR shows that while DRL can efficiently explore chemical space, current practices in dataset selection, split strategy, and reward design do not by themselves guarantee fair outcomes across disease areas, scaffolds, or physicochemical profiles.
At the same time, fairness cannot be reduced to a single proxy score or fully delegated to automated pipelines; domain expertise and ethical judgment remain essential. Future work should therefore focus on developing fairness-aware benchmarks with scaffold- and indication-stratified splits. It should also prioritize parity metrics defined on disease areas and chemotypes. In addition, clear reporting standards are needed to surface trade-offs between efficacy, safety, and parity. In doing so, DRL systems can better support balanced and transparent data-driven strategies in modern drug discovery.

\section{Abbreviations and Acronyms}

\begin{table}[h]
\caption{List of abbreviations used in this study.}
%  \vspace{-1mm}
\label{tab:abbreviations}
\renewcommand{\arraystretch}{1.05}
\small
\centering
\begin{tabular}{p{0.18\columnwidth}p{0.72\columnwidth}}
\hline
\textbf{Abbrev.} & \textbf{Definition} \\
\hline
MW      & Molecular weight \\
logP    & Lipophilicity     \\
HBD     & Hydrogen-bond donor \\
HBA     & Hydrogen-bond acceptor \\
QED     & Quantitative Estimate of Drug-likeness \\
DRL     & Deep reinforcement learning \\
SAS     & Synthetic accessibility score \\
DRD2    & Dopamine Receptor D2 \\
MolDQN  & Molecule Deep Q-Networks \\
QADD    & Quality assessment-based drug design approach \\
GCPN    & Graph Convolutional Policy Network \\
CNS     & Central nervous system \\
ADMET   & Absorption, distribution, metabolism, excretion, and toxicity \\
TPSA    & Topological polar surface area \\
RGFN    & Reaction-GFlowNet \\
\hline
\end{tabular}
 % \vspace{-2mm}
\end{table}

Table~\ref{tab:abbreviations} summarizes the abbreviations and technical terms frequently used throughout this paper.

% -------------------------------------------------------------------------
\printbibliography
\end{document}